\newcommand{\mrvir}{r_{vir}}
\newcommand{\rvir}{$\mrvir$}

\newcommand{\hMpc}{\ensuremath{h^{-1}\,\mathrm{Mpc}}}

\newcommand{\dll}{d_\mathrm{LL,\perp}}
\newcommand{\dpar}{d_\mathrm{LL,\parallel}}

\newcommand{\zfrb}{\ensuremath{z_{\rm FRB}}}
\newcommand{\zfg}{\ensuremath{z_{\rm fg}}}

\newcommand{\dmunits}{\ensuremath{\rm pc \, cm^{-3}}} 
\newcommand{\dmihalo}{\ensuremath{\mathrm{DM}_\mathrm{halo}}}
\newcommand{\sigdmhalo}{\ensuremath{\sigma(\dmihalo)}}
\newcommand{\sigdmhalos}{\ensuremath{\sigma(\dmhalos)}}
\newcommand{\dmhalo}{\ensuremath{\mathrm{DM}_\mathrm{halo}}}
\newcommand{\dmhalos}{\ensuremath{\mathrm{DM}_\mathrm{halos}}}
\newcommand{\dmahalos}{\ensuremath{\langle \dmhalos \rangle}}

\newcommand{\dmaigm}{\ensuremath{\langle {\rm DM}_{\rm IGM} \rangle}}

\newcommand{\dmfrb}{\ensuremath{\mathrm{DM}_\mathrm{FRB}}}
\newcommand{\dmigm}{\ensuremath{\mathrm{DM}_\mathrm{IGM}}}

\newcommand{\dmhost}{\ensuremath{\mathrm{DM}_\mathrm{host}}}
\newcommand{\avgdmhost}{\ensuremath{\overline{\dmhost}}}
\newcommand{\dmmw}{\ensuremath{{\rm DM}_{\rm MW}}}
\newcommand{\dmmwhalo}{\ensuremath{{\rm DM}_{\rm MW,halo}}}
\newcommand{\dmmwism}{\ensuremath{{\rm DM}_{\rm MW,ISM}}}

\newcommand{\dmacosmic}{\ensuremath{\langle {\rm DM}_{\rm cosmic} \rangle}}
\newcommand{\dmcosmic}{\ensuremath{\mathrm{DM}_\mathrm{cosmic}}}
\newcommand{\dmecosmic}{\ensuremath{\dmcosmic^{\rm est}}}
\newcommand{\dmmcosmic}{\ensuremath{\dmcosmic^{\rm model}}}
\newcommand{\sigdmmcosmic}{\ensuremath{\sigma(\dmmcosmic)}}

\newcommand{\rperp}{\ensuremath{R_\perp}}

\newcommand{\pdmcosmic}
{\ensuremath{p(\dmcosmic|z)}}

\newcommand{\mhalo}{\ensuremath{M_\mathrm{halo}}}
\newcommand{\mstar}{\ensuremath{M_\mathrm{*}}}

\newcommand{\msun}{\ensuremath{{\rm M}_\odot}}

\newcommand{\fohseven}{FRB20190714A}
\newcommand{\fohnine}{FRB20200906A}
\newcommand{\fohfour}{FRB20200430A}
\newcommand{\fohone}{FRB20210117A}

\documentclass[twocolumn]{aastex631}
\usepackage{amsmath}
\usepackage{graphicx}
\usepackage[normalem]{ulem}
\usepackage{natbib}

\shorttitle{Excess FRB DM}
\shortauthors{Simha et al.}
\graphicspath{{./}{figures/}}

\begin{document}

\title{Searching for the sources of excess extragalactic dispersion of FRBs}

\correspondingauthor{Sunil Simha}
\email{shassans@ucsc.edu}


\author[0000-0003-3801-1496]{Sunil Simha}
\affil{University of California - Santa Cruz
1156 High St.
Santa Cruz, CA, USA 95064}

\author[0000-0001-9299-5719]{Khee-Gan Lee}
\affil{Kavli IPMU (WPI), UTIAS, The University of Tokyo, Kashiwa, Chiba 277-8583, Japan}

\author[0000-0002-7738-6875]{J. Xavier Prochaska}
\affil{University of California - Santa Cruz
1156 High St.
Santa Cruz, CA, USA 95064}
\affil{Kavli IPMU (WPI), UTIAS, The University of Tokyo, Kashiwa, Chiba 277-8583, Japan}
\affil{Division of Science, National Astronomical Observatory of Japan,
2-21-1 Osawa, Mitaka, Tokyo 181-8588, Japan}

\author[0000-0003-0574-7421]{Ilya S. Khrykin}
\affil{Kavli IPMU (WPI), UTIAS, The University of Tokyo, Kashiwa, Chiba 277-8583, Japan}

\author[0000-0002-0298-8898]{Yuxin Huang}
\affil{Kavli IPMU (WPI), UTIAS, The University of Tokyo, Kashiwa, Chiba 277-8583, Japan}

\author[0000-0002-1883-4252]{Nicolas Tejos}
\affil{Instituto de F\'isica, Pontificia Universidad Cat\'olica de Valpara\'iso, Casilla 4059, Valpara\'iso, Chile}

\author[0000-0003-1483-0147]{Lachlan Marnoch}
\affiliation{School of Mathematical and Physical Sciences, Macquarie University, NSW 2109, Australia}
\affil{CSIRO, Space and Astronomy, PO Box 76, Epping NSW 1710 Australia}
\affil{Astronomy, Astrophysics and Astrophotonics Research Centre, Macquarie University, Sydney, NSW 2109, Australia}
\affil{ARC Centre of Excellence for All-Sky Astrophysics in 3 Dimensions (ASTRO 3D), Australia}

\author[0000-0002-5934-9018]{Metin Ata}
\affiliation{
The Oskar Klein Centre, Department of Physics, Stockholm University,  \\ AlbaNova University Centre, SE 106 91 Stockholm, Sweden
}

\author[0009-0002-9608-9275]{Lucas Bernales}
\affiliation{Instituto de F\'isica, Pontificia Universidad Cat\'olica de Valpara\'iso, Casilla 4059, Valpara\'iso, Chile}

\author[0000-0003-3460-506X]{Shivani Bhandari}\thanks{Veni fellow}
\affil{ASTRON, Netherlands Institute for Radio Astronomy, Oude Hoogeveensedijk 4, 7991 PD
Dwingeloo, The Netherlands}
\affil{Joint institute for VLBI ERIC, 
Oude Hoogeveensedijk 4, 7991 PD Dwingeloo, The Netherlands}

\author[0000-0001-5703-2108]{Jeff Cooke}
\affil{Centre for Astrophysics and Supercomputing, Swinburne University of Technology, Mail Number H29, PO Box 218, 31122, Hawthorn, VIC, Australia}
\affil{ARC Centre of Excellence for All-Sky Astrophysics in 3 Dimensions (ASTRO 3D), Australia}
\affil{CSIRO, Space and Astronomy, PO Box 76, Epping NSW 1710 Australia}

\author[0000-0001-9434-3837]{Adam T.Deller}
\affiliation{Centre for Astrophysics and Supercomputing, Swinburne University of Technology, Mail Number H29, PO Box 218, 31122, Hawthorn, VIC, Australia}

\author[0000-0003-4501-8100]{Stuart D. Ryder}
\affil{School of Mathematical and Physical Sciences, Macquarie University, NSW 2109, Australia}
\affil{Astronomy, Astrophysics and Astrophotonics Research Centre, Macquarie University, Sydney, NSW 2109, Australia}

\author[0000-0001-5310-4186]{Jielai Zhang}
\affil{Centre for Astrophysics and Supercomputing, Swinburne University of Technology, Mail Number H29, PO Box 218, 31122, Hawthorn, VIC, Australia}
\affil{ARC Centre of Excellence for Gravitational Wave Discovery (OzGrav), Australia}


\begin{abstract}
The FLIMFLAM survey is collecting spectroscopic data of field galaxies near fast radio burst (FRB) sightlines to constrain key parameters describing the distribution of matter in the Universe. In this work, we leverage the survey data to determine the source of the excess extragalactic dispersion measure (DM), compared to the Macquart relation estimate of four FRBs: \fohseven, \fohnine, \fohfour, and \fohone. By modeling the gas distribution around the foreground galaxy halos and galaxy groups of the sightlines, we estimate \dmhalos, their contribution to the FRB dispersion measures. The \fohseven\ sightline shows a clear excess of foreground halos which contribute roughly 2/3$^{rd}$ of the observed excess DM, thus implying a sightline that is baryon-dense. \fohnine\ shows a smaller but non-negligible foreground halo contribution, and further analysis of the IGM is necessary to ascertain the true cosmic contribution to its DM. \fohfour\ and \fohone\ show negligible foreground contributions, implying a large host galaxy excess and/or progenitor environment excess.

\end{abstract}



\section{Introduction} \label{sec:intro}

With the advent of the concordance $Lambda$-Cold Dark Matter ($\Lambda$~CDM) cosmological paradigm, there is now a comprehensive model for the large-scale structure of matter in the universe, and its formation under the influence of gravity is one of the key tests that is actively being researched. Cosmic microwave background (CMB) experiments \citep[e.g.][]{wmap2013,planck2018} have precisely measured the contents of the universe and simulations have rendered clarity regading the time-evolution of structure beginning from primordial fluctuations \citep[e.g.][]{Springel+05}. In the current paradigm, dark matter forms the cosmic web, the large scale structure that includes voids, filaments, and dense halos and serves as scaffolding for the accretion of baryonic matter. 
Indeed, hydrodynamical simulations \citep[e.g.][]{tngbaryons, eaglegas, horizonsim} have shown us that the ionized gas populates dark matter halos and also occupies the cosmic web filaments or the intergalactic medium (IGM), albeit in a much more diffuse state.

The low density of the IGM plasma has long challenged baryon census studies at $z\lesssim0.5$. The Lyman alpha forest and UV absorption studies of metal ion tracers such as OVI and OVII are not sensitive to $\sim40\%$ of the IGM baryons \citep[i.e. the Missing-Baryon Problem;][]{Fukugita+98,Shull+2012} which reside is the hot ($\sim10^6~K$), diffuse phase according to theory \citep[e.g.][]{cen+ostriker2006}. With existing facilities, very long-exposure X-ray observations  (multi-million seconds) are required to to detect the weak absorption expected from OVII tracers of the hot phase \citep[e.g.][]{Nicastro+18}. Alternatively, stacking the weak kinetic Sunyaev-Zeldovich signal between $\gtrsim10^6$ galaxy pairs could reveal the gas in filaments \citep{sz_effect_filaments}.

In the meantime, the serendipitous discovery of the first Fast Radio Burst (FRB) in archival data \citep{lorimer2007} has set in motion a series of paradigm-changing discoveries. FRBs are millisecond-duration radio transients whose origins are still widely debated. With improved radio detection techniques, over the last five years multiple FRBs have been localized in the sky with sub-arcsecond accuracy \citep{tendulkar+2017,Bannister+19,law+2020,chime_181030} and thus their radial distance could be confidently measured from their host galaxy redshifts (\zfrb). FRBs pulses are dispersed by plasma during propagation and the extent of this effect is directly related to the integrated, line-of-sight free electron density ($n_e$). 
This effect is quantified by the FRB Dispersion Measure (\dmfrb) which is defined as:
    \begin{equation}
         \dmfrb = \int \frac{n_e}{1+z}dl \;\;\;.
    \end{equation}
Here, $z$ is the cosmological redshift and $dl$ is the distance element along the line-of-sight. 
As \dmfrb~is an integral quantity, 
it may be represented as the sum 
of the electron reservoirs encountered 
during propagation. i.e. 
\begin{equation}
    \dmfrb = \dmmw+\dmcosmic+\dmhost .
\end{equation}
Here, \dmmw{} is from the electrons within the Milky Way interstellar medium (ISM) and halo, \dmhost{} is from the counterpart structures in the host galaxy, and \dmcosmic{} is from the plasma in intervening halos and the diffuse IGM in the foreground, i.e. $\dmcosmic = \dmhalos+\dmigm$. \citet{Macquart+2020} were the first to estimate \dmcosmic{} for a sample of localized FRBs and showed that it is correlated with \zfrb. This was as expected of the current paradigm of cosmological expansion and the fraction of ionized baryons in the universe \footnote{Estimated by leveraging observational constraints on denser baryon reservoirs in the form of stars, remnants and neutral gas \citep[e.g.][]{Fukugita04, Macquart+2020}.}. This proved directly that the ``Missing'' Baryons were not just found, but also that \dmfrb~could viably probe the diffuse plasma in the Universe. The community has largely adopted the moniker of the ``Macquart relation" to refer to the average \dmcosmic, i.e. \dmacosmic, versus \zfrb.

While the mean Macquart relation is well described by 
cosmology \citep[e.g.][]{inoue04}, 
there is expected to be scatter about \dmcosmic\
at any given redshift
due to the inhomgeneity of cosmic structure.
For example, some FRB sightlines 
may intersect the gas rich environments of intra galaxy cluster media while others may primarily intersect cosmic voids. 
Furthermore, galaxy feedback can influence the variance in gas density by distributing gas further out of gravitational wells 
\citep[e.g.][]{xyz19}. Indeed, as we shall show in the subsequent section, one identifies a number of FRBs where estimates for \dmcosmic~from nominal assumptions on \dmhost~imply $\dmcosmic > \dmacosmic$. However, it is not evident \textit{a priori} if the excess arises from foreground structure (i.e. intervening halos and IGM overdensities) or from an atypical host and progenitor environment. Our previous work \citep{Simha+2020,Simha+2021} has introduced
a methdology to estimate the contribution from foreground halos. Here, we apply our analysis to four FRB sightlines with apparently high \dmcosmic~values. Future application of such analyses on a statistical sample of FRBs
can inform us on the distribution of ionized gas within dark matter halos \citep[e.g.][]{McQuinn2014,xyz19,Lee+2022}.

To this end, we leverage the redshifts of galaxies collected as part of the FRB Line-of-Sight Ionization Measurement From Lightcone AAOmega Mapping  (FLIMFLAM) survey \citep{Lee+2022}. This redshift survey aims to study the foreground matter distribution along $\sim30$ FRB sightlines. The key results expected from the survey include constraints accurate to $\sim10\%$ on (1) the fraction of baryons in the universe in the diffuse IGM and (2) the fraction of baryons residing in circum-galactic halos that are in the ionized phase. In this redshift survey, spectroscopic redshifts 
 and photometry of foreground galaxies within $\sim1$ degree of an FRB sightline are used to generate bespoke models of the line-of-sight ionized matter density tailored to individual lines-of-sight, which can then be compared with the DM from the FRB. Key reservoirs of said matter include intervening dark matter halos and the diffuse intergalactic medium (IGM). In this work, with a subset of the spectroscopic data collected, we investigate four excess \dmcosmic~sightlines: \fohseven, \fohfour, \fohnine\ and \fohone. These fields were targeted with the wide-field Anglo-Australian Telescope(AAT)/AAOmega and the Keck/LRIS and DEIMOS spectrographs.

This manuscript is outlined as follows: section \ref{sec:data} describes the data collection and reduction. Section \ref{sec:dm_halo} describes our intervening-galaxy-halo DM estimation procedure. Section \ref{sec:results} describes the results and section \ref{sec:discussion} discusses their implications. Throughout this work, unless otherwise specified, we assume a $\Lambda$CDM cosmology with Planck 2018 cosmological parameters \citep{planck2018}.

\section{Data} \label{sec:data}

\subsection{Sample selection}
\begin{figure*}[ht!]
    \centering
    \includegraphics[width=\textwidth]{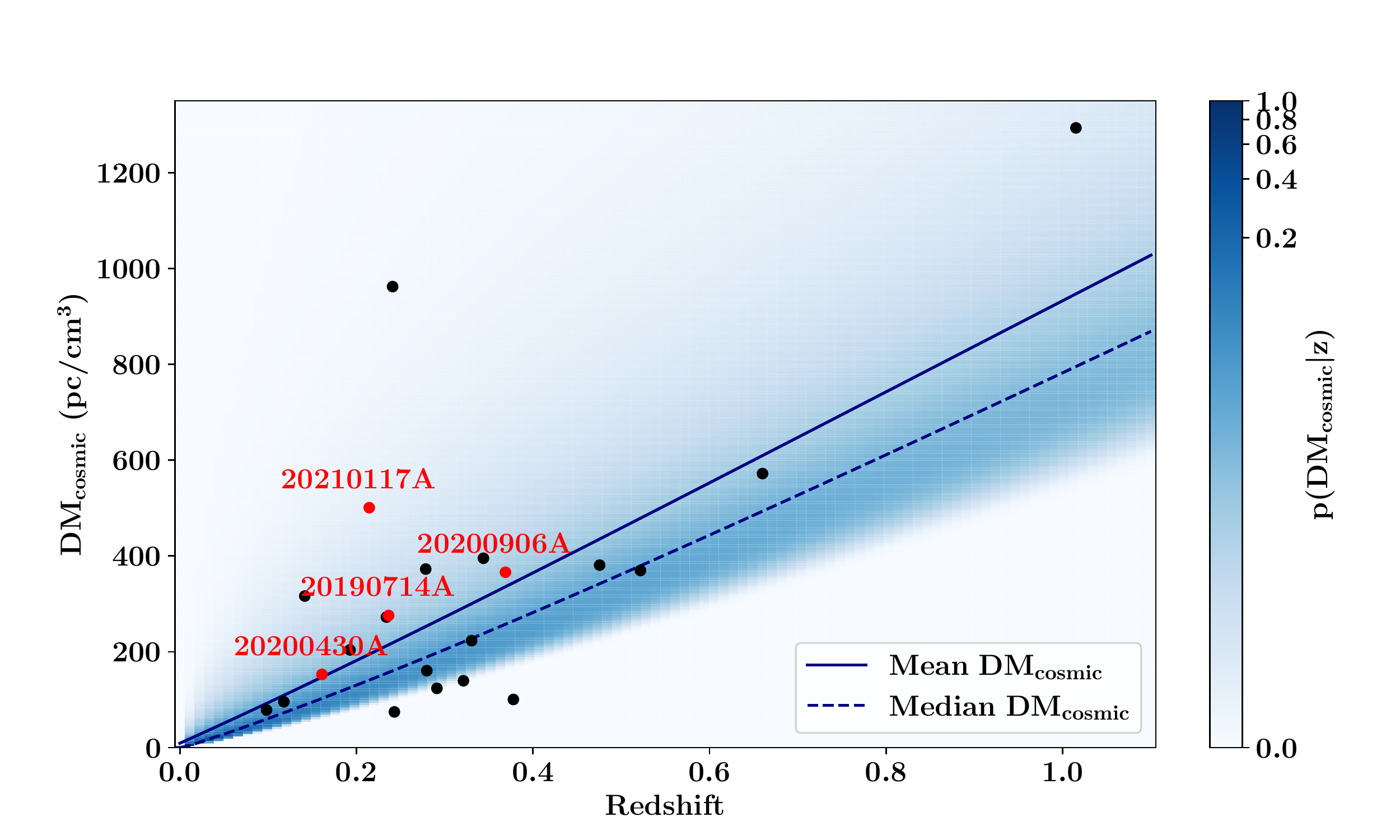}
    \caption{An updated Macquart relation plot including published well-localized FRBs from CRAFT at $z<0.7$. The solid line is the mean 
    \dmacosmic\ from a universe with the
    $\Lambda$CDM cosmology, a.k.a. the Macquart relation. The blue shading represents 
    \pdmcosmic, the PDF of \dmcosmic~at each redshift given the variance in the matter density along a random sightline in the universe from intervening halos and the gas in the cosmic web filaments. Note the median of the distribution (dashed line) lies lower than the mean, implying that most sightlines are expected to have few intervening foreground halos that contribute significantly
    to \dmcosmic. The data points are 
    estimates \dmecosmic\ for FRBs from the
    CRAFT survey.  These are the observed \dmfrb\  corrected for the Milky Way contribution 
    and an assumed host contribution of 
    $\avgdmhost = 186~\dmunits$ 
    in the rest frame. The sightlines examined in this work are marked in red, all of which have $\dmecosmic>\dmacosmic$. Of the other notably high \dmecosmic~sources, FRB20190520B ($\zfrb\sim0.23$) at $\sim1000$ \dmunits~will be analyzed in a future work.}
    \label{fig:macquart}
\end{figure*}

As described in the introduction, structure
in the cosmic web is expected to produce a 
significant scatter in the Macquart relation due to sightline-to-sightline variation in the column density of intervening gas \citep{Macquart+2020}. 
Figure~\ref{fig:macquart} is an updated plot showing the 
Macquart relation and data from the sample of CRAFT-localized FRBs published to date \citep{Macquart+19_sample, Bhandari+2019_sample, Qiu+2019, James+2022_sample}. 
The DM values shown in the plot correspond to 
estimates of the cosmic dispersion measures,

\begin{equation}
\dmecosmic = \dmfrb - \dmmw - \avgdmhost/(1+z),
\label{eqn:dmcosmic_est}
\end{equation}
where \dmmw~is estimated as the sum of the ISM contribution (\dmmwism) 
taken from the NE2001 model \citep{Cordes2003}, and the halo contribution (\dmmwhalo) which is assumed 
to be 40~\dmunits. We do note that there is evidence pointing to highly variable Milky Way halo contribution, \dmmwhalo. i.e. $\sigma(\dmmwhalo)\sim 100~\dmunits$. For example, \citet{Das+2021} use X-ray absorption lines in quasar spectra from gas within the Milky Way CGM and constrain \dmmwhalo~along numerous sightlines. Though we did not find a matching absorption sightline from their dataset within 3 degrees of our FRBs we acknowledge the possibility of large \dmmwhalo. Studies such as \citet{cook+23} and \citet{ravi+23} involving low \dmfrb~sightlines ($\lesssim 100~\dmunits$) place tighter constraints ($\dmmwhalo=28-111~\dmunits$). In this context, we concede our assumption for \dmmwhalo~is probably low but has little impact on our qualitative findings. 
Furthermore, for Figure~\ref{fig:macquart},
we assume a median host 
contribution of $\avgdmhost = 186~\dmunits$. 
A primary goal of this paper is to distinguish between
these two scenarios, i.e. the excess arising from the foreground or the FRB host, along individual sightlines.

The blue shading visualizes the expected probability density of \dmcosmic\ at each redshift, 
\pdmcosmic, with an assumed feedback parameter $F=0.31$ \citep{Macquart+2020, McQuinn2014}. 
The long, low-probability tail
in \pdmcosmic\ to high \dmcosmic\ values is due to
massive halos of galaxy clusters and groups, which occasionally intersect a sightline. One sees that a sizable fraction of the FRB sample lies
above the Macquart relation, and a subset have \dmecosmic~values 
at or beyond the 80th percentile of the expected distribution at their redshifts. Naively, assuming that our ansatz for \avgdmhost is correct, one would expect only 20\% (i.e. $\sim4$) of the sightlines on average above the 80th percentile for the sample size shown in the figure. However, we find 11.

The FRBs with $\dmecosmic > \dmacosmic$ may
arise from higher host contributions than the assumed average
(i.e. $\dmhost > \avgdmhost$) 
or a larger than average foreground contribution to
\dmcosmic, or both. 
Of the 11 FRBs with this apparent excess in \dmcosmic, 
6 have been targeted in the FLIMFLAM survey and have both shallow, wide-field ($m_r<20$ mag within 1.1 deg radius around the FRB) AAT/AAOmega  spectroscopy, plus deeper, narrow-field spectra ($m_r<23$ $within \sim5$ arcmin radius) using the Keck/LRIS and Keck/DEIMOS instruments. One field, FRB20190608A was previously studied by \citet{Simha+2020} using redshift data from SDSS and KCWI integral-field unit observations. In a separate paper, we will use a slightly different methodology to analyze the foreground contribution to the well-studied high-DM source FRB20190520B (Lee et al., in prep). In this work, we present the foreground analysis of the other four fields: \fohseven, \fohfour, \fohnine\ and \fohone. 
All of these have \dmecosmic\ near or beyond the 
80th percentile in \pdmcosmic\ as listed in
Table~\ref{tab:excessDM}.

\begin{table*}
    \centering \hspace*{-1in}
    \caption{\footnotesize Our sample \label{tab:excessDM}}
    \begin{tabular}{|c|c|c|c|c|c|c|c|}
    \hline
    FRB & RA & Dec & Redshift & \dmfrb & \dmacosmic & \dmecosmic & Percentile \\
     & $\deg$ & $\deg$ &  & \dmunits & \dmunits & \dmunits &  \\
    \hline
    FRB20190714A & 183.97971 & -13.02100 & 0.2365 & 504.1 & 205 & 275 & 88 \\
    FRB20200430A & 229.70642 & 12.37675 & 0.1610 & 380.0 & 137 & 152 & 81 \\
    FRB20200906A & 53.49617 & -14.08318 & 0.3688 & 577.8 & 326 & 366 & 82 \\
    FRB20210117A & 339.97929 & -16.15142 & 0.2145 & 731.0 & 185 & 502 & 97 \\
    \hline
    \end{tabular}
    Notes: \dmacosmic\ is the mean \dmcosmic\ at the FRB redshift.
     \dmecosmic\ is the estimated \dmcosmic\ value for the FRB based on \dmfrb\ and an assumed $\avgdmhost= \, 186~\dmunits$.
     Percentile is the percentage of FRBs expected to have $\dmcosmic < \dmecosmic$ at the FRB redshift.
    \end{table*}

\subsection{Spectroscopic target selection}

\begin{figure*}
    \centering
    \includegraphics[width=\textwidth]{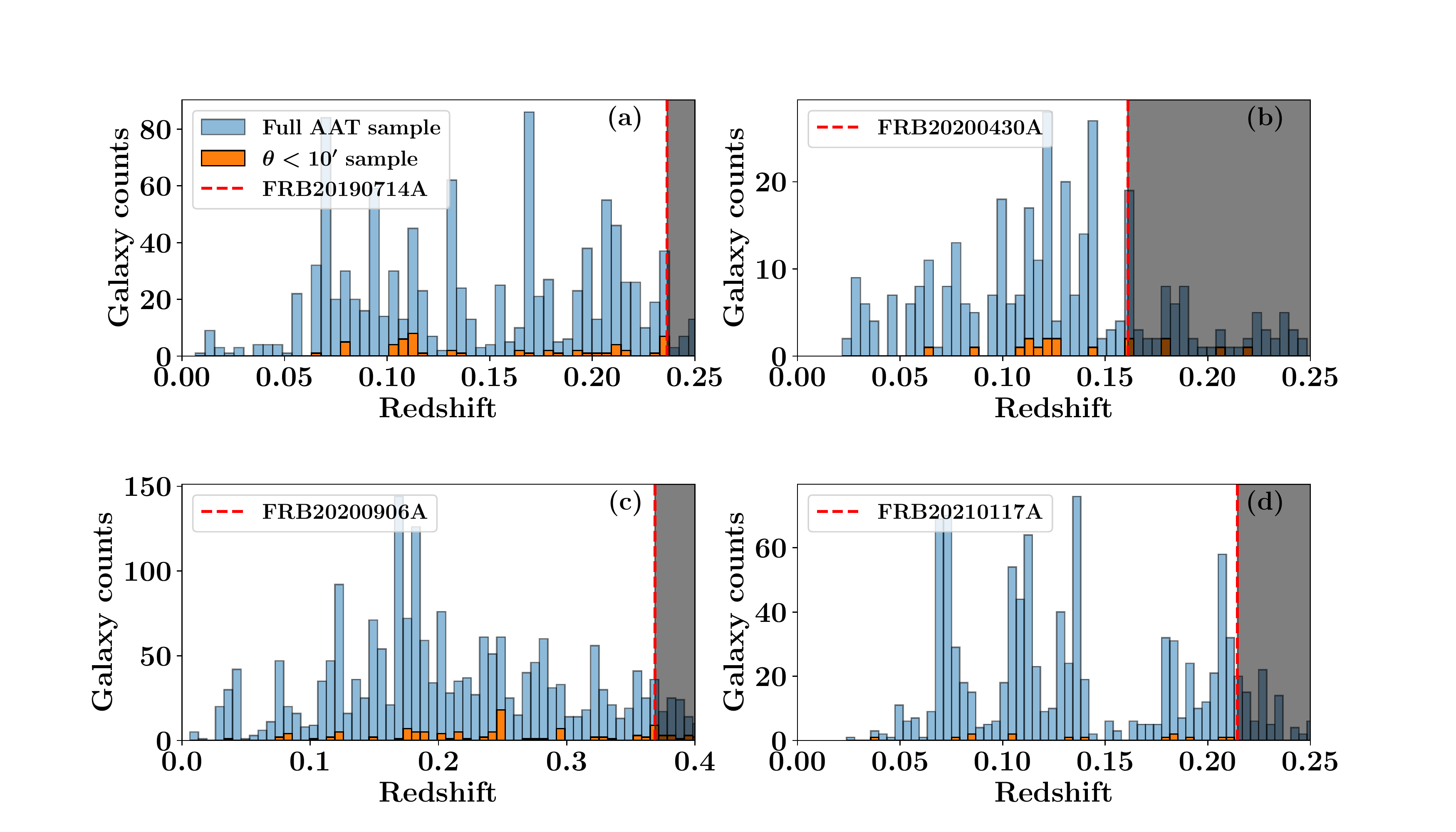}
    \caption{Histogram of galaxy redshifts obtained from the AAOmega spectrograph in the four fields. The full 1.1 degree radius sample is shown in blue and the subset of galaxies within 10 arcmin is shown in orange. The FRB redshift is marked by the dashed red line, and the shaded region represents background galaxies that are not relevant to this study}
    \label{fig:aat_hist}
\end{figure*}

Field galaxies within a radius of 1.1 degrees of the sightlines were targeted using the fiber-fed AAOmega spectrograph on the 3.9m Anglo-Australian Telescope (AAT) at Siding Spring, Australia. For two fields (\fohseven~and \fohone), the fiber configurations were designed to target sources with $m_r<19.4$ mag that were well-resolved in the Pan-STARRS imaging, i.e. distinct from point sources. For fields \fohfour and \fohnine, the target criterion is $m_r<19.2$ mag and $m_r<19.8$ mag respectively that were well-resolved in DECam imaging from archival DESI Legacy Imaging Surveys data \citep{Dey2019}. Due to unfavorable weather conditions, we were unable to observe the full roster of fiber configurations generated for \fohfour, and so this field has sparser wide-field coverage than intended. We therefore supplement our spectroscopic data on this field from the SDSS database. Each fiber configuration was observed for  $\sim1$ hr in the 1x1 binning mode with the 570 nm dichroic, which split the light into red and blue components. The red camera used the 385R grating blazed at 720nm while the blue camera used the 580V grating and the blaze is set to 485nm. The red and blue spectra were reduced, coadded and combined using the 2dFDR version 6.2 based on python 2.7 kindly provided by the OzDES group \citep{10.1093/mnras/stv1507,Childress_2017}. 
We used the MARZ \citep{MARZ} software to determine redshifts, which cross-correlates the input spectra with a set of templates and determines the best redshift. This was followed by a visual inspection to confirm the redshifts, with adjustments as necessary.
Figure \ref{fig:aat_hist} shows the histogram of redshifts obtained from the AAT for the fields analyzed in this paper. The spectroscopic success rate of the survey, which is defined by the fraction of the number of targets with secure redshift and the total number of the targets that were observed, is around 90\%.

In addition, the FRB fields were targeted with the Keck DEIMOS and LRIS spectrographs in the multi-object spectroscopy mode. 
 
We used Pan-STARRS $r$-band imaging to select $m_r<23$ mag galaxies (i.e. as before, rejecting point sources) within $\sim5$ arcmin of the sightline. To further limit sources to $z\lesssim0.3$, we rejected sources that satisfy these color criteria based on our analysis of mock galaxy photometry \citep{Lee+2022}:
\begin{equation}
    \begin{aligned}
        g-r & >0 \\
        r-i & >0.7 \\
        i & >20.5
    \end{aligned}
\end{equation}

With LRIS, multi-object slitmask-based spectroscopy of the target galaxies was performed. Our configuration was as follows: 600/7500 grating for the red-side, 600/4000 grism for the blue side and the 560D dichroic. All raw frames were binned 2x2. The LRIS observations were obtained only for the fields of \fohseven\ and \fohfour\ during a previous run and not all objects in the field could be covered due to limited time. The galaxies that were omitted were subsequently targeted with DEIMOS. All LRIS/DEIMOS spectra were reduced with v1.2 of the PypeIt package \citep{PypeIt} package. We set a detection threshold of 3$\sigma$ above the noise floor for object identification and forced detection for fainter objects using the slitmask information stored in the metadata of the raw frames. Our DEIMOS observations were obtained on a later run with the 600ZD grating and GG455 order blocking filter and 1x1 binning. Each mask configuration was observed for $\sim50$ min. Together, 95\% of the candidate galaxies within 5 arcmin of the FRB were targeted.

We ignored the serendipitous spectra, i.e. spectra of non-targeted sources captured in our slits, as they generally had no discernible features for redshift assignment. We did not flux-calibrate the spectra as this is not necessary for redshift estimation from line features.

As with the AAT spectra, all reduced spectra from Keck were processed via MARZ \citep{MARZ} to determine redshifts, followed by a visual inspection. 
As with the AAT data, of the targeted Keck spectra, $>90\%$ had good redshift assignments.

In the case of \fohseven, \citet[][in prep.,]{Marnoch+2023} have presented MUSE IFU pointing of $0.67$\,hours with the Wide Field Mode (WFM) covering the $1\arcmin \times1\arcmin$ area around the FRB sightline. Of the 61 galaxies extracted from the stacked white light image (i.e. the image averaged over the spectral dimension), 7 were identified to be foreground sources.

The reduced spectra with their assigned redshifts are made available as a specDB \citep{specDB} file \footnote{Available at this \href{https://drive.google.com/drive/folders/1gVVwTAmiHJtFB_GD56hgmDylXHYbOTFS?usp=sharing}{Google Drive link}}. This is an HDF5 file with the convenience functions from the specDB package for easy retrieval of data.

\subsection{Photometric data}
\label{sec:photom}
To estimate foreground galaxy properties such as stellar mass, we fit the publicly available flux measurements with a spectral energy distribution (SED) model. To this end, we used the \emph{grizy} photometry from the Pan-STARRS \citep{pan-starrs} catalog, \emph{W1}, \emph{W2}, \emph{W3}, and \emph{W4} from the WISE All-Sky source catalog \citep{wise} and supplemented with the \emph{YJHKs} photometry from the Vista Hemisphere Survey (VHS) catalog \citep{vista} where available. The details regarding the SED fitting procedure are elucidated in the following section.

\section{DM halo analysis} 
\label{sec:dm_halo}

In this section, we describe the methodology
implemented to estimate the dispersion measure
contributed from the halo of a galaxy or
group of galaxies, \dmihalo.  
We refer to the summed quantity
along a given sightline as \dmhalos.

\subsection{Individual Halos}

Once spectroscopic redshifts were assigned, the available photometry was fit with an SED using CIGALE \citep{cigale}. We assumed a delayed-exponential star-formation history with no burst population, a synthetic stellar population prescribed by \citet{bc03}, the \citet{Chabrier03} initial mass function (IMF), dust attenuation models from \citet{calzetti01}, and dust emission templates from \citet{dale14}, where the AGN fraction was capped at 20\%. This provided an estimate of the stellar mass, \mstar, of the 
foreground galaxy at a given redshift \zfg.

We then translate \mstar~to galactic halo mass, \mhalo, using the mean stellar to halo mass relation (SHMR) described by \citet{moster+13} at that \zfg. Subsequently, \dmhalo~were estimated using the \citet{xyz19} modified NFW halo profile model. We assumed that the total amount of baryons in the halo trace the cosmic mean ($\Omega_b/\Omega_m$).
We assumed the halo gas extends to one virial radius 
(\rvir) and that 75\% of the baryons are in the hot, ionized phase in the halo. This assumes that 25\% of the baryons in the galaxy is in condensed forms \citep[e.g. stars and neutral gas; see][]{Fukugita+98}. While this fraction
may vary with halo properties \citep[e.g.][]{Behroozi+10} or assumptions on galaxy feedback \citep{Sorini2022,Ayromlou2022}, we emphasize that this is a relatively conservative maximal model for the
CGM of galaxies, i.e. one may consider the DM estimates as upper limits.
Adopting this CGM model, we then integrate the dispersion
measure of the gas at the observed 
impact parameter \rperp\ of the galaxy from the 
sightline determined from its redshift \zfg\ and 
the angular offset.

The uncertainties in the \mstar~estimation and the SHMR relation propagate into the \dmhalos~estimate. For each galaxy, we assumed that the $\log\mstar$ distribution at a given redshift was Gaussian with the mean and standard deviations obtained from CIGALE. Accounting for the error in the SHMR is more involved as it depends on both \mstar~and galaxy redshift. The SHMR is described in Equation 2 of \citet{moster+13} with 8 parameters. We took the best fit parameters and uncertainties from their Table~1 as the mean and standard deviations of the independent normal distributions that these parameters were sampled from. We ignored any co-variance in these fit parameters. From the $\log\mstar$ distributions, 1000 samples are drawn and the SHMR parameter space is sampled 1000 times for each $\log\mstar$ realization. Thus, for every galaxy, we produce $10^6$ $\log\mhalo$ realizations, and subsequently, \dmhalo\ estimates. The mean and variance from these individual distributions are used when drawing our conclusions for the sightlines. 

\subsection{Galaxy group contributions}
\label{sec:groups}
It is important to account for galaxy groups or clusters, since the overall halo mass is typically much larger than the sum of the putative member masses if estimated individually.
This results in DM contributions much
greater than that estimated for individual
group members.
To search for galaxy groups within the FLIMFLAM spectroscopic catalog, we make use of an anisotropic friends-of-friends (FoF) group finder that has previously been applied to SDSS galaxy survey data (\citealt{tago:2008}; but see also \citealt{tempel+2012},\citealt{tempel+2014}). 
This finder assumes a transverse linking length, $\dll$, which varies as a function of redshift, $z$, in the following way:

\begin{equation}
\dll(z) = d_\mathrm{LL,0}[1 + a \,\mathrm{arctan}(z/z_*)],
\end{equation}
where $d_\mathrm{LL,0}$ is the linking length at the initial redshift, whereas
$a$ and $z_*$ are parameters governing the redshift evolution. 
This redshift-dependent linking length allows one, in principle, to account for the 
declining completeness of the galaxies with increasing redshift in a flux-limited spectroscopic survey.
The line-of-sight linking length, $\dpar$, is then set as a fixed multiple of $\dll$; 
the ratio $\dpar/\dll$ is another free parameter for the group finder. 
To determine the appropriate values for these free parameters, 
we ran the group finder on the FLIMFLAM catalogs and manually iterated the free parameters of the group finder, while
 visually inspecting the resulting groups from the FLIMFLAM catalog in both the transverse and line-of-sight
dimensions at each iteration.  
Our criteria was to ensure the selection is not so permissive as to include cosmic web filament structures as part of the identified groups, while simultaneously not being so stringent as to omit the more massive groups at the high-redshift end where the data is typically sparser. We arrived at the following values for the group-finding in this paper:
$\dll=0.2\,\hMpc$, $a=0.75$, $z_*=0.1$, and $\dpar/\dll=10$.  

To limit ourselves to reasonably robust groups, we select for a minimum richness of $N_\mathrm{gal}\geq 5$. Furthermore, we apply the same modified NFW profile model; limited still to one virial radius but scaled up to the group mass estimated as our fiducial model. In addition to the coordinates and redshift of each group center, the code also provides a halo mass estimate by applying the virial theorem on the projected group radius and velocity dispersion\footnote{The group catalogs generated for our fields are available at this \href{https://drive.google.com/drive/folders/1gVVwTAmiHJtFB_GD56hgmDylXHYbOTFS?usp=sharing}{Google Drive link}.}.

\subsection{Halo Contributions}
\label{sec:avg_dmhalos}

While our analysis can provide estimates of \dmhalos~for individual sightlines, it is useful to compare them against a mean cosmic contribution from halos for any random sightline up to \zfrb. One may produce a theoretical estimate of this as follows:

Adopting the halo mass function (HMF) (using the implementation of \citet{HMF}) and restricting ourselves to $\mhalo<10^{16}~\msun$, we can estimate the total number of halos of each mass bin expected to intersect within 1~\rvir~of each sightline. Using our baryon distribution model described in section \ref{sec:dm_halo}, this can be translated to the average \dmhalos~along the sightline, i.e. \dmahalos. 

\begin{figure}
    \centering
    \includegraphics[width=0.5\textwidth]{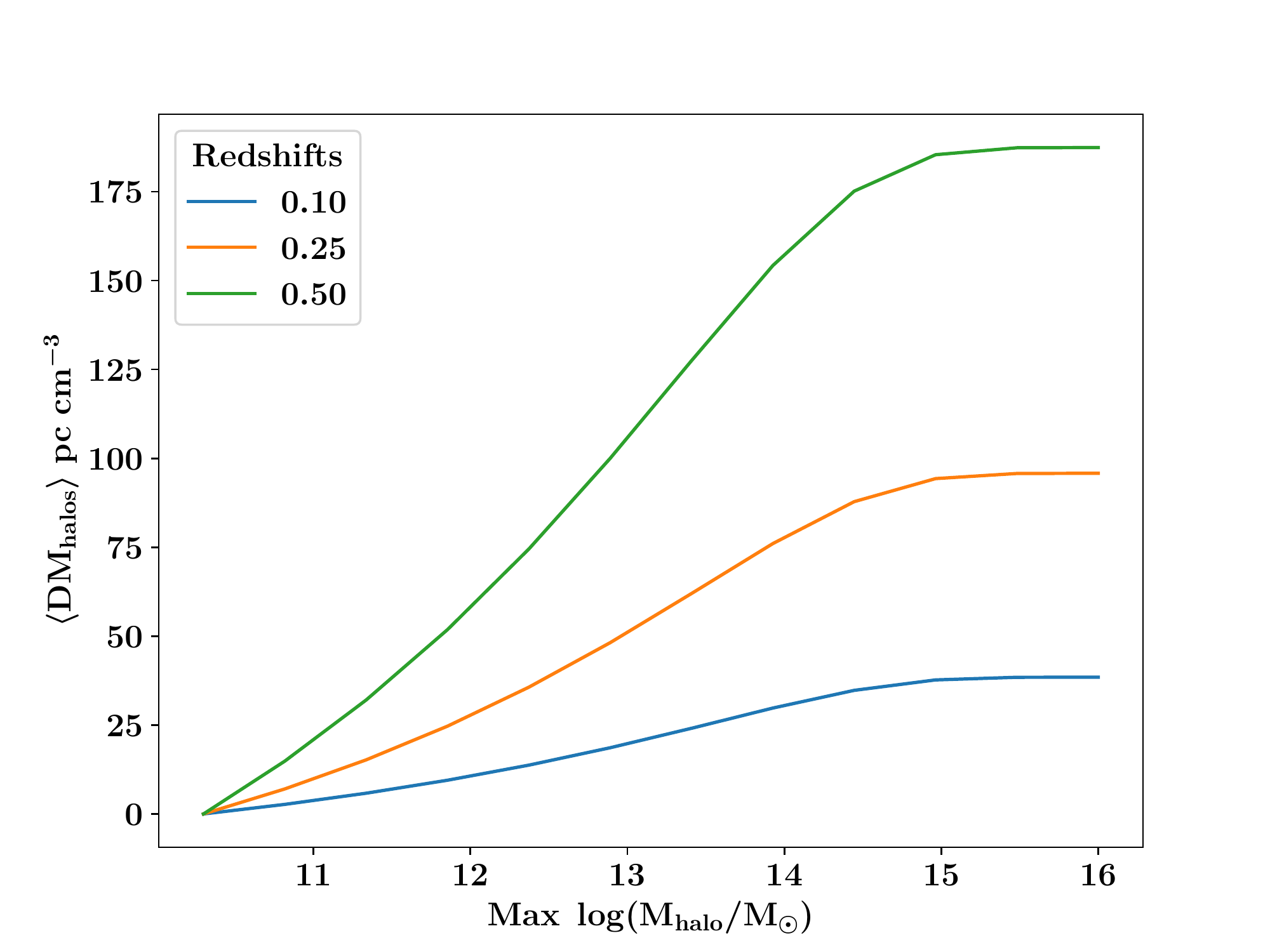}
    \caption{Cumulative estimate of \dmahalos\ as a function of the maximum halo mass that can contribute to \dmhalos.
    \dmahalos~is computed assuming the halo mass function corresponding to our adopted cosmology \citep{HMF}, integrated to the given maximum 
    \mhalo\ from the same minimum $\mhalo= 10^{10.3} \, \msun$. The halo gas model has the same modified NFW profile described previously, extending to one virial radius with 75\% of the halo baryons in the hot, ionized phase.}
    \label{fig:dmhalosavg}
\end{figure}

This average represents an upper limit as we are considering halos more massive than those of galaxies. \dmahalos\ monotonically increases with the halo mass up to which the HMF is integrated over (see Figure \ref{fig:dmhalosavg}) but plateaus near $\mhalo \approx 10^{15}\msun$. 
This presumably reflects the low average probability of intersecting such massive, but rare, halos.
Changing the model parameters that influence \dmhalo\ 
have similar effect on \dmahalos. e.g. increasing the assumed fraction of ionized baryons in the halo scales up both \dmhalo\ and \dmahalos\ by the same factor.

\begin{figure*}[ht!]
    \centering
    \includegraphics[width=1.1\textwidth,trim={1in 2.5in 1in 2.5in},clip]{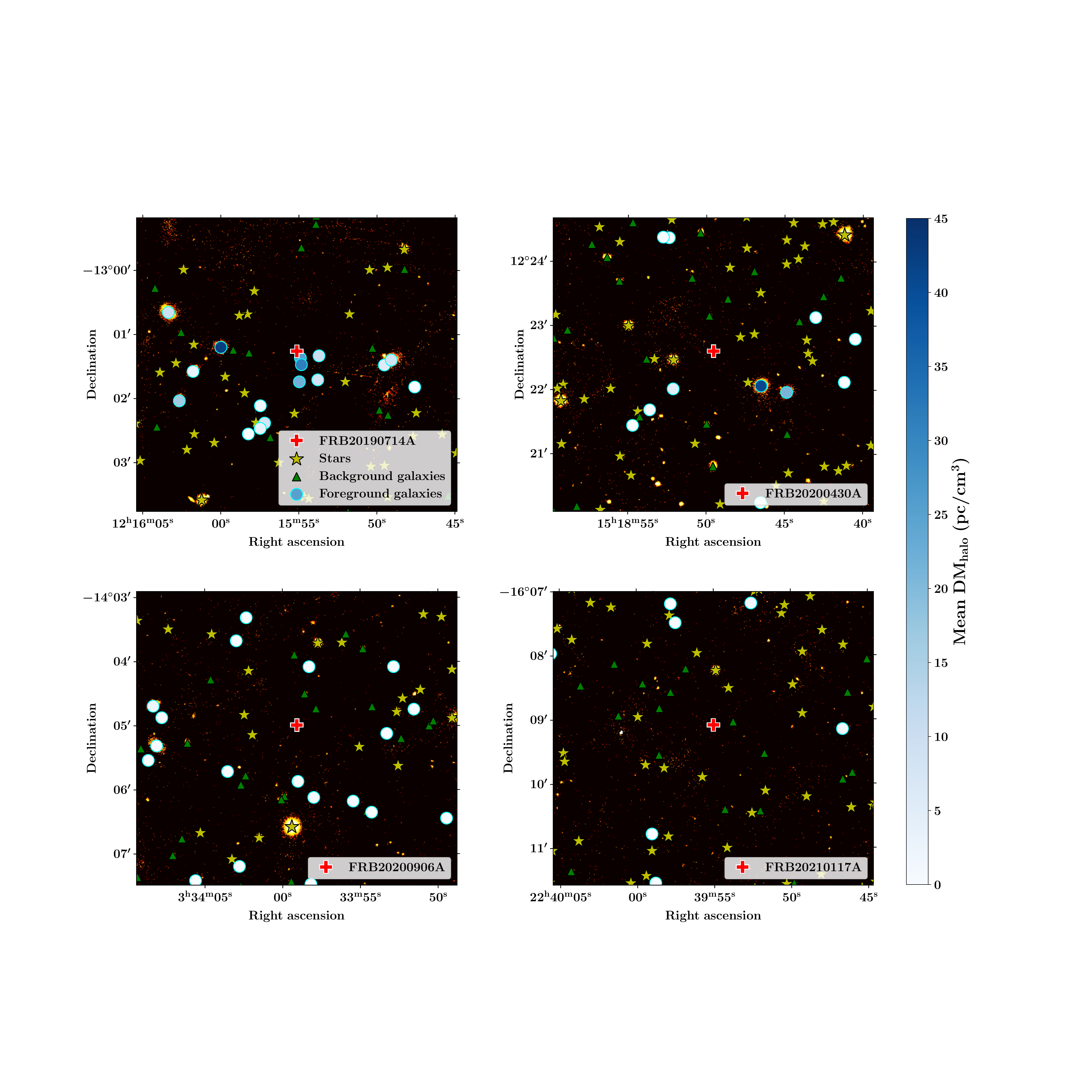}
    \caption{\footnotesize 
    Zoomed-in ($5' \times 5'$)  illustration of the
    fields and results for the four
    FRB sightlines: (a) \fohseven, (b) \fohfour, (c) \fohnine, and (d) \fohone. The background shows Pan-STARRS r-band images. In each image, the red crosses mark the location of the FRB, the green triangles mark the background galaxies and the yellow stars mark the point sources that were ignored from spectroscopic targeting. The blue circles mark the foreground galaxies, with the color scaled 
    according to the estimated \dmihalo\ value. 
    }
    \label{fig:fg_halos_sky1}
\end{figure*}

\section{Results}
\label{sec:results}

The analysis described above was applied to each galaxy in each field, resulting in probability distributions for the \dmhalos~contribution of  individual galaxies and groups. The \dmhalos\ value is then the straight sum along each 
sightline. Our findings from the analysis for each sightline described above are presented in this section. 

Figure \ref{fig:fg_halos_sky1} is a visual summary of the individual fields. It highlights stars, background objects and foreground objects $\lesssim3$ arcmin of the FRBs on the 
$r$-band image of the field from Pan-STARRS. The foreground objects are colored by the average \dmhalo~contribution estimated for each of them.

\begin{figure*}[ht!]
    \centering
    \includegraphics[width=\textwidth]{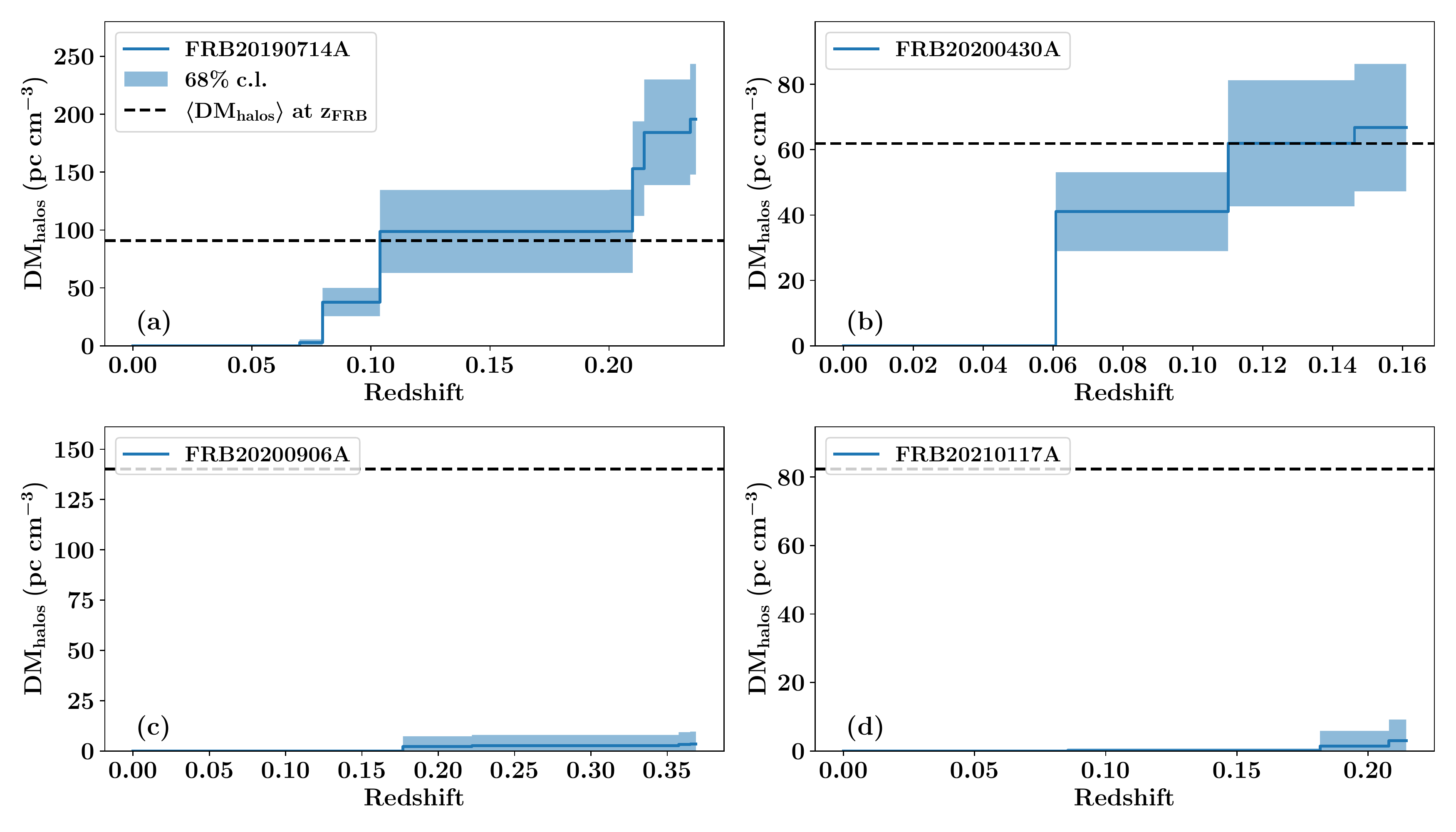}
    \caption{Empirical evaluation of \dmhalos\ for the FRB sightlines as a function of redshift. 
    The blue curve presents the cumulative 
    estimation of \dmhalos\ 
    from $z = 0$, which increases monotonically
    as foreground halos are encountered along the sightline.
    The blue shading represents 68\% confidence limits on the \dmhalos\ estimate, which is the running quadrature sum of the individual 1-sigma limits of the DM distributions for the individual galaxies. The black dashed lines represent estimates for \dmahalos~assuming the our adopted halo mass function (up to $\mhalo = 10^{16}~\msun$) and the adopted halo 
    gas distribution model used to calculate \dmhalos. 
    While the \fohseven\ sightline clearly exceeds the average expectation, both \fohone\ and \fohnine\ are barely in excess of 
    $\dmhalos = 0 \, \dmunits$. 
    \fohfour\ exhibits a \dmhalos\ value 
    consistent with \dmahalos.}
    \label{fig:dm_cumsum}
\end{figure*}

\begin{table*}
\centering \hspace*{-1in}
\caption{\footnotesize Foreground galaxies contributing to \dmhalos.\label{tab:dmhalos}}
\begin{tabular}{|c|c|c|c|c|c|c|c|c|}
\hline
FRB & RA & Dec & \zfg & \rperp & $\rm \log(\mstar/\msun)$ & $\rm \log(\mhalo/\msun)$ & \dmihalo & \sigdmhalo \\
 & $\deg$ & $\deg$ &  & kpc &  &  & \dmunits & \dmunits \\
\hline
FRB20190714A & 184.01105 & -13.03391 & 0.1044 & 236 & 10.6 & 12.2 & 16.6 & 16.3 \\
FRB20190714A & 183.97902 & -13.02895 & 0.2119 & 102 & 10.1 & 11.7 & 21.9 & 10.4 \\
FRB20190714A & 183.97876 & -13.02276 & 0.0802 & 11 & 8.2 & 10.7 & 25.1 & 4.2 \\
FRB20190714A & 183.97849 & -13.02450 & 0.2141 & 47 & 9.6 & 11.4 & 31.4 & 20.4 \\
FRB20190714A & 183.99997 & -13.02000 & 0.1042 & 140 & 10.8 & 12.4 & 42.7 & 29.5 \\ \hline
FRB20200430A & 229.71715 & 12.36691 & 0.1448 & 135 & 9.7 & 11.4 & 4.8 & 2.9 \\
FRB20200430A & 229.68695 & 12.36605 & 0.1109 & 163 & 10.5 & 11.9 & 21.0 & 15.0 \\
FRB20200430A & 229.69376 & 12.36773 & 0.0619 & 67 & 10.3 & 11.8 & 41.0 & 12.1 \\ \hline
FRB20200906A & 53.53467 & -14.07823 & 0.1761 & 417 & 11.0 & 13.1 & 2.2 & 5.1 \\ \hline
FRB20210117A & 339.96901 & -16.11978 & 0.1827 & 378 & 10.5 & 12.0 & 1.4 & 4.4 \\
FRB20210117A & 339.94438 & -16.15251 & 0.2085 & 424 & 10.9 & 12.8 & 1.6 & 4.2 \\
\hline
\end{tabular}
\end{table*}

\subsection{\fohseven}

Examining Figure~\ref{fig:fg_halos_sky1}, one
notes multiple galaxies in the foreground field
of \fohseven\ including several within 
$\approx 30''$.  These galaxies lie primarily
at two redshifts: $z= 0.10$ and 0.21 and have
estimated halo masses that yield significant
\dmhalo\ contributions.
The galaxy with the smallest impact parameter 
(J121554.90-130121.95)
was found in the VLT/MUSE datacube 
and has a redshift of 0.08 yielding a 
projected perpendicular distance of 
$\rperp = 11$\,kpc \citep[][in prep.]{Marnoch+2023}. 
Even though its mass estimate indicates it is a dwarf galaxy ($\mstar = 10^{8.5}~\msun$), its close proximity to the sightline leads to a substantial \dmhalo\ contribution of $25 \, \dmunits$. 

The wide-field data from Keck/DEIMOS and LRIS show 110 foreground galaxies, and of these 17 show non-zero \dmhalo\ contributions. 
Table \ref{tab:dmhalos} lists the foreground galaxies and their mean \dmhalo\ contributions \footnote{The full galaxy catalogs with their halo masses and \dmhalo~estimates for our fields are available at this \href{https://drive.google.com/drive/folders/1gVVwTAmiHJtFB_GD56hgmDylXHYbOTFS?usp=sharing}{Google Drive link}.}.  

We do not find any group contribution when applying our fiducial halo gas model, which truncates at the virial radius, to the groups identified in this field.
If, however, one extended the model
to two virial radii we estimate one of the groups 
would give a $50 \,\dmunits$ contribution. 
This group is centered at RA/Dec of (184.1382405, $-$13.0107427) and $z=0.111$. The FRB sightline is at a transverse distance of 1.16 Mpc. With 20 member galaxies and a halo mass of $10^{13.9} \msun$, this group may potentially contribute to \dmcosmic. We do not include this contribution in our \dmhalos~estimate but discuss the implications of doing so in Section~\ref{sec:discussion}.

Figure~\ref{fig:dm_cumsum}a presents the
cumulative sum of \dmhalos\ with redshift
and shows a total value of
$200 \pm 45 \, \dmunits$.
This exceeds by over $100 \, \dmunits$ 
the average estimated \dmahalos\ 
for the FRB redshift using the 
methodology described in Section~\ref{sec:avg_dmhalos}.
For this FRB, we infer that its 
\dmecosmic\ exceeds \dmacosmic\ 
owing to an excess of foreground structure.
We return to this conclusion in the following
section.

\subsection{\fohfour}
While \fohfour~has the least significant excess value of \dmecosmic~in our sample, we estimate that  
the foreground galaxies in the field of \fohfour\ 
contribute significantly to \dmhalos, similar to \fohseven.
Specifically, we 
estimate $\dmhalos = 65 \pm 20 \, \dmunits$ which 
is 
comparable to \dmahalos\ at $\zfrb=0.161$ (Fig. \ref{fig:dm_cumsum}b). 

We do not find any group contribution to \dmhalos~for this sightline; the closest group lies at
4.6\,Mpc transverse distance with a mass of only $10^{13} \, \msun$. 
At over $\sim$10 virial radii from the sightline, this group has no plausible
influence on the observed $\dmhalos$.

\subsection{\fohnine}

Although this field exhibits a large number of
foreground galaxies within $10'$ of the FRB 
including nearly 20 within $5'$ of the sightline,   
we estimate their contributions \dmcosmic\ 
to be nearly negligible. 
Many of these galaxies also have high, estimated
halo masses but their individual contributions
are generally $\dmihalo \lesssim 1~\dmunits$
(Table~\ref{tab:dmhalos}).
This results from the
the large physical impact parameters;
only one has $\rperp < 200$\,kpc 
from the sightline.

We estimate no group contribution to \dmhalos\ for 
this field, with the closest group being 860 kpc away with a mass of $10^{11.7}\msun$ ($z=0.04$).
This comparatively low-mass halo was detected as a group only by virtue of its low redshift (and hence small distance modulus).

\subsection{\fohone}

From our sample of four FRBs with 
$\dmecosmic > \dmacosmic$, 
\fohone\ is the most extreme outlier with more than $380 \, \dmunits$ in excess of the average value at the \zfrb~= 0.2145. 
Remarkably, as is evident from Figure~\ref{fig:fg_halos_sky1}b, 
we do not find any foreground halos in close
proximity to the sightline.
As such, 
the total \dmhalos\ estimate is very 
small 
(Figure \ref{fig:dm_cumsum}b). 
The galaxy with the largest \dmihalo\ estimate
(1.6 \dmunits) is over $400$ kpc away and has a halo mass of $10^{12.8} \msun$. 
Given the uncertainties in halo masses,
\dmhalos\ is even consistent with 0, 
i.e.\ no intersections within one virial
radius of any foreground halo.

Not surprisingly,
we also find no contribution from the galaxy groups identified in this field.
The closest group lies at a distance of 2\,Mpc.

\begin{figure*}[ht!]
    \centering
    \includegraphics[width=\textwidth]{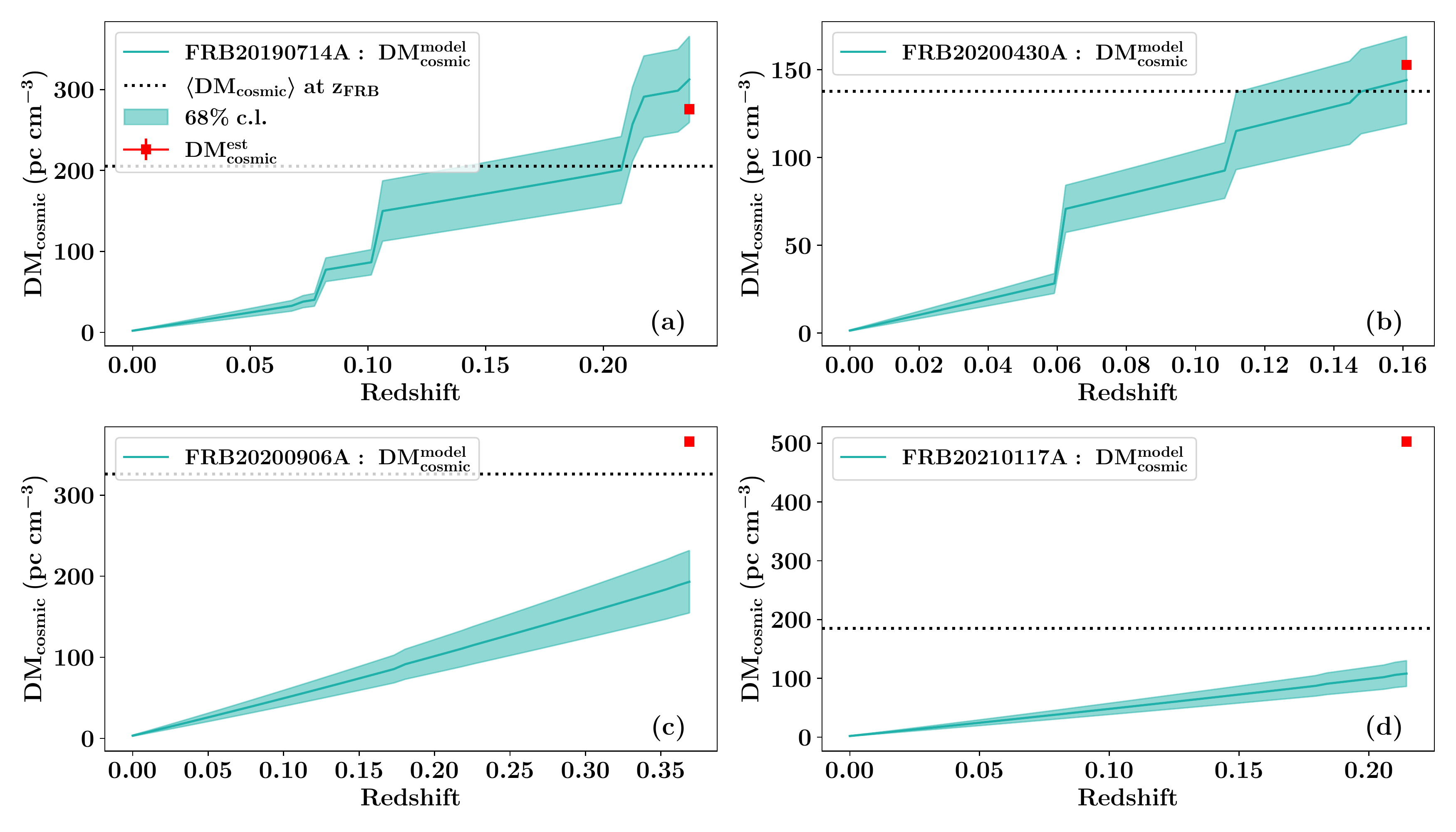}
    \caption{Estimates of \dmcosmic~for the FRB sightlines as a function of redshift. The solid, teal curve is \dmmcosmic, a sum of \dmhalos, i.e. the solid, blue curve from Figure \ref{fig:dm_cumsum} and an estimate of the average
    IGM contribution to \dmcosmic\ (see text for details).
    The dotted line shows \dmacosmic\ at \zfrb~in each subplot. The shading around the solid curve represents a 68\% confidence limit which includes an assumed 20\% uncertainty for \dmaigm~in quadrature with the uncertainties from Figure \ref{fig:dm_cumsum}. The red point is an estimate of \dmcosmic~for each FRB taken from Figure \ref{fig:macquart}, i.e. by subtracting the assumed
    host and Milky Way contributions.}
    \label{fig:dm_cumsum_igm}
\end{figure*}

\section{Discussion}
\label{sec:discussion}

In the previous section we presented our analysis of the foreground matter distribution along four sightlines, with focus on \dmhalos. We now discuss the implications of these results. The primary motivation of this paper was to explore the origin of apparent excess in \dmcosmic\ along FRB sightlines.
To place our results in this context, we construct an
empirical model \dmmcosmic\ for the four sightlines
based on our findings.  Specifically, define

\begin{equation}
\dmmcosmic = \dmhalos + \dmaigm
\end{equation}
where \dmaigm\ is given by

\begin{equation}
\dmaigm = \dmacosmic - \dmahalos
\label{eqn:avg_igm}
\end{equation}
with \dmahalos\ calculated as described in 
Section~\ref{sec:avg_dmhalos} and all quantities
are evaluated at \zfrb.
In future analyses the FLIMFLAM survey will
estimate \dmigm\ for individual fields with
the cosmic web reconstruction algorithm ARGO \citep{argo},
which is a Bayesian estimator for the matter density field given the foreground galaxy halo masses and 3D locations (i.e. their sky position and redshifts). 

Our \dmmcosmic\ estimate
assumes 
the uncertainty in \dmhalos\ (Table~\ref{tab:summary})
and a $20\%$ statistical uncertainty in \dmaigm\
based on numerical simulations \citep[e.g.][]{Lee+2022}
We also emphasize that the assumed CGM model used to 
estimate \dmhalos\ impacts \dmahalos\ and therefore
\dmaigm\ through Equation~\ref{eqn:avg_igm}.
This sensitivity to the CGM model (here a systematic error)
lies central to correlating \dmfrb\ against
galactic halos and large-scale structure to constrain
properties of halo gas and the baryonic content of the
IGM \citep{Lee+2022, CHIME_cross_corr}.
In the current analysis, however, the CGM model
has less impact for decreases in \dmhalos\ will
be compensated by an increase in \dmaigm.

Figure~\ref{fig:dm_cumsum_igm} presents cumulative
estimates for \dmmcosmic\ with redshift 
for each field.  These are compared with 
\dmacosmic\ at \zfrb\ and our values
for \dmecosmic\ using Equation~\ref{eqn:dmcosmic_est}.
As one may have anticipated, \dmmcosmic\
for the two fields with large \dmhalos\ values
(\fohseven, \fohfour) are consistent with
\dmecosmic\ (see also Table~\ref{tab:summary}).
For these two FRBs, we have empirical confirmation
of the theoretical paradigm for \dmcosmic, i.e.\ that
its intrinsic scatter tracks the incidence of
foreground structure.  These results lend further
confidence for future analyses leveraging \dmfrb\ 
to resolve the cosmic web. Compared to the previously studied sightlines of FRB20190608A \citep{Simha+2020} and FRB20180924B \citep{Simha+2021}, \fohseven~is the first that shows a significantly large contribution from foreground halos. As mentioned previously, such sightlines are expected to be rare \citep[e.g.][]{McQuinn2014}.

On the other hand,  the \dmmcosmic\ estimates for
\fohnine\ and \fohone\ do not even meet the average
\dmacosmic\ for these sources, much less the apparent
excess implied by \dmecosmic.  
The shortfalls are $\approx 200\,\dmunits$
and $\approx 425\,\dmunits$ respectively.
Even accounting for uncertainty in our \dmhalos\ and
\dmigm\ estimates, one cannot account for these differences
within the $\sim~1\sigma$ uncertainties.
This suggests the observed excess is due to a higher
than average \dmhost\ component; we estimate
the rest-frame \dmhost\ values to be
$\approx 422\,\dmunits$
and $\approx 665\,\dmunits$ respectively. These sightlines are remarkably similar to the that reported by \citet{Niu+2022} for FRB20190520B, implying a relatively low \dmcosmic~compared to \dmhost{}
for these sightlines. Future detection of such sightlines might be key to unravel the likely progenitor scenarios and in investigating how \dmhost~depends on host galaxy properties.

We may further assess the likelihood of this conclusion
as follows.  Adopting a lognormal PDF for \dmhost\ 
with the parameters estimated by 
\cite{James+2022},
the fraction of FRBs with \dmhost\ values in 
excess of these estimates are 18\%\ and 9\%\ 
respectively. The large \dmhost~values can be attributed to a combination of the local progenitor environment and the host ISM. 

One may search for signatures of a high \dmhost\ value
from detailed studies of the host galaxies.
\fohone\ arises in a low-mass (dwarf) galaxy with a 
low star-formation rate \citep{Bhandari+2023,gordon+2023}.
It is offset from the galaxy center by $\approx 3$\,kpc
which exceeds the half-light radius.
In these regards, there is nothing apparent in the host
properties nor its inferred halo that would suggest
such a large \dmhost\ value. \citet[][in prep.]{Bhandari+2023} propose a possible scenario involving the FRB progenitor being embedded in the outflows of a hyper-accreting black hole and note that long-term, short-cadence observations of the FRB polarization may constrain such a model should the FRB repeat.

\fohnine~on the other hand arises from a high-mass, high-star-formation-rate galaxy and is coincident with the disk of the host \citep[see Figure 1 of ][]{gordon+2023}. This implies a fraction of the \dmhost~arises from the host ISM. For example, \citet{chittidi+21} estimated for FRB20190608B $\sim90~\dmunits$ for the host ISM contribution from the local H-alpha line emission measure. While \citet{gordon+2023} report slightly lower star-formation rate for the host of \fohnine~than FRB20190608B, one can visually discern a higher disk inclination for the former and speculate a comparable if not higher \dmhost\ 
for the ISM component. A dedicated optical follow-up study of the host with an integral-field unit, especially if one can resolve $\lesssim 1 \rm kpc$ around the FRB, can help place upper limits on the ISM contribution. For all the four FRB host galaxies, if we applied our galaxy halo gas model and computed \dmhalo\ as analyzed above, we estimate a contribution of $\lesssim35~\dmunits$ each.

As mentioned previously, a full IGM reconstruction analysis is necessary for a complete understanding of the foreground matter density, e.g. as done for FRB20190608B \citep{Simha+2020}. While we have established two of our fields have $\dmmcosmic\sim\dmecosmic$, it is possible that the IGM reconstruction may reveal $\dmigm>\dmaigm$ and therefore lay tighter constraints on \dmhost. With $\sim30$ sightlines, the FLIMFLAM survey will perform such analysis and render, as a useful by-product, a posterior distribution for \dmhost. This distribution can serve as a prior to future FRB-based IGM tomography work as well as to constrain FRB progenitor channels.

\section{Conclusion}
To summarize, we analyzed the galaxies in the foreground
of four localized FRBs, whose estimated cosmic dispersion measure
\dmecosmic\ significantly exceeds the average at \zfrb.
Implementing the methodology detailed in Section~\ref{sec:dm_halo},
we estimated the DM contribution of foreground galactic and group halos, \dmhalos, as summarized in Table~\ref{tab:summary}.
For two fields, we found a high incidence of halos
at close impact parameters to the sightline, such that
the \dmhalos\ estimate matches or exceeds the average cosmic
expectation value, \dmahalos.
For the other two fields, the \dmhalos\ estimate
is less than 5\,\dmunits\ owing to the absence
of foreground halos near the sightline. Our results reinforce the paradigm that FRBs can effectively probe foreground matter overdensities. That being said, one must exercise caution in accounting for plasma in the host galaxy and immediate FRB progenitor environment when studying matter distribution
along the sightline. Combined with \citet{Simha+2020}
we conclude FRBs with apparent high \dmcosmic\ 
arise from both higher than average foreground structure
and inferred higher host contributions, with nearly
equal probability.

Thus the FLIMFLAM survey is ramping up efforts towards data collection and analysis. Future results are expected to lay robust constraints on the parameters describing foreground matter distribution as well as constrain \dmhost~statistically.

\section*{Acknowledgements}
We thank Elmo Tempel for kindly providing his group-finding software, and Chris Lidman for assistance with the
OzDES data reduction pipeline for AAOmega.
Authors S.S., J.X.P.,  and N.T., 
as members of the Fast and Fortunate for FRB
Follow-up team, acknowledge support from 
NSF grants AST-1911140, AST-1910471
and AST-2206490. N.T. and L.B. acknowledge support by FONDECYT grant 11191217.
We acknowledge generous financial support from Kavli IPMU that made
FLIMFLAM possible.
Kavli IPMU is supported by World Premier International Research Center Initiative (WPI), MEXT, Japan.
Based on data acquired at the Anglo-Australian Telescope, under programs A/2020B/04, A/2021A/13, and O/2021A/3001. We acknowledge the traditional custodians of the land on which the AAT stands, the Gamilaraay people, and pay our respects to elders past and present.

\software{
MARZ \citep{MARZ},
HMFEmulator \citep{HMF},
specDB \citep{specDB},
CIGALE \citep{cigale},
Astropy \citep{astropy:2018},
Numpy \citep{numpy},
Scipy \citep{scipy},
Matplotlib \citep{Hunter:2007}.
}

The Python scripts used to perform our analysis are available in our FRB GitHub repository (https://github.com/FRBs/FRB). 

\begin{table*}
\centering \hspace*{-1in}
\caption{\footnotesize Summary Table \label{tab:summary}}
\begin{tabular}{|c|c|c|c|c|c|c|c|}
\hline
FRB & Redshift & \dmahalos & \dmhalos & \sigdmhalos & \dmecosmic & \dmmcosmic & \sigdmmcosmic \\
 &  & \dmunits & \dmunits & \dmunits & \dmunits & \dmunits & \dmunits \\
\hline
FRB20190714A & 0.2365 & 92 & 195 & 47 & 275 & 312 & 53 \\
FRB20200430A & 0.1610 & 63 & 66 & 19 & 152 & 144 & 24 \\
FRB20200906A & 0.3688 & 142 & 3 & 4 & 366 & 193 & 38 \\
FRB20210117A & 0.2145 & 83 & 3 & 4 & 502 & 108 & 21 \\
\hline
\end{tabular}
\end{table*}


\bibliography{bibliography.bib}{}
\bibliographystyle{aasjournal}



\end{document}